\documentclass[pra,floatfix,twocolumn,showpacs,amsmath,amssymb]{revtex4}
\usepackage{graphicx}
\usepackage{dcolumn}
\usepackage{bm}
\usepackage{color}
\usepackage{ulem}

\begin{document}
\title{
Observable precursor of topological phase transition: temperature-dependent electronic specific heat in two-dimensional Dirac fermions
}

\author
{Keita Kishigi}
\affiliation{Faculty of Education, Kumamoto University, Kurokami 2-40-1, 
Kumamoto, 860-8555, Japan}

\author
{Keita Tsukidashi}
\affiliation{Faculty of Advanced Science and Technology, Kumamoto University,  Kurokami, 2-39-1, Kumamoto, 860-8555, Japan}


\author{Yasumasa Hasegawa}
\affiliation{Department of Material Science, 
Graduate School of Science, 
University of Hyogo, Hyogo, 678-1297, Japan}

\date{\today}

\begin{abstract}
Dirac points in two-dimensional massless Dirac fermions 
are topologically protected. Although single Dirac point cannot 
disappear solely, a pair of two Dirac points annihilates after merging at a time-reversal invariant momentum (TRIM). This process triggers a topological phase transition. In this paper, we numerically calculate the electronic specific heat ($C$) of the systems with a honeycomb lattice and $\alpha$-(BEDT-TTF)$_2$I$_3$ in the case that two Dirac points are moving and merging by changing the ratio of the magnitude between the transfer integrals, which can be controlled by uniaxial pressure for example. 
When two Dirac points are close to but not at TRIM, the temperature dependence exhibits a crossover from $C\propto T^{2.0}$ (expected for separated Dirac points) at low temperatures ($T\leq T_{\rm co}$) to $C\propto T^{1.5}$ (expected in the case of the merged Dirac points) 
at high temperatures ($T\geq T_{\rm co}$). Here, $T_{\rm co}$ denotes the crossover temperature, which is determined by the potential barrier magnitude between two Dirac points. Our findings demonstrate that the precursor of the topological phase transition is observed through the temperature-dependence of the electronic specific heat.
\end{abstract}

\date{\today}

\maketitle

\section{Introduction}

Graphene is a well-known two-dimensional system consisting of massless Dirac fermions\cite{novo2005}, where the Dirac points originating from Dirac equation in relativistic quantum mechanics exist. The energy dispersion at the Dirac points is linear. These Dirac points are topologically protected, ensuring their stability even under the application of uniaxial pressure, except for pair annihilation.

In the realm of quasi-two-dimensional organic conductors\cite{review,review2}, the presence of a two dimensional Dirac fermion system has been unveiled in $\alpha$-(BEDT-TTF)$_2$I$_3$ through tight-binding model analysis\cite{Katayama2006} and the 
first-principles band calculations\cite{Kino,Alemany2012}, and the existence of the Dirac point at the Fermi energy ($\varepsilon_{\rm F}$) has been also experimentally confirmed through the analysis of transport quantities\cite{Kajita2014,Hirata2011,Osada2008} and thermodynamic quantities such as electronic specific heat\cite{Konoike2012}.  
Similarly, for $\alpha$-(BETS)$_2$I$_3$ it has been recently revealed  via first-principle calculations\cite{Ohki2020,Tsumu2021,kitou2021} and through magnetoresistance measurements\cite{Tajima2021}. 

From recent theoretical proposals\cite{morinari2020} and the experiment\cite{tajima2023} under a magnetic field, it has been suggested that $\alpha$-(BEDT-TTF)$_2$I$_3$ is a three-dimensional Dirac fermion. However, for the study of electronic specific heat ($C$), we expect that its system can still be approximated as a two-dimensional system at temperatures ($T$) above several Kelvin due to the relatively small interlayer coupling ($t_{\perp}\sim$ 1 meV)\cite{tajima2023}. In this study, we neglect the quasi-two-dimensionality in $\alpha$-(BEDT-TTF)$_2$I$_3$ when studying electronic specific heat. In fact, a power law behavior, $T^{1.8}$, has been observed, particularly below 4 K\cite{Konoike2012}. This temperature dependence of the electronic specific heat is close to that in 
two-dimensional Dirac systems, $C \propto T^{2.0}$.


Theoretical studies have demonstrated the movement and merging of two Dirac points at time-reversal invariant momentum (TRIM) in two-dimensional systems with honeycomb lattice\cite{Dietl2008,Hasegawa2006}, VO$_2$/TiO$_2$ nanostructures\cite{Bane}, and quasi-two-dimensional systems like $\alpha$-(BEDT-TTF)$_2$I$_3$\cite{Suzumura2013}. Merging Dirac points are referred to as semi-Dirac points\cite{Bane}. After merging, a pair of Dirac points annihilates, leading to a topological phase transition to a normal insulator. The existence of semi-Dirac points and the occurrence of topological phase transitions have been observed in artificial systems such as ultracold atoms on optical lattices\cite{Tarruell2012}, lattices of photonic resonators\cite{Bellec}, and polaritons in lattices of semiconductor micropillars\cite{Mil,real}.

The energy dispersion at the semi-Dirac point is linear in two directions (e.g., $\pm k_x$-directions) and quadratic in other two directions ($\pm k_y$-directions) \cite{Dietl2008,Hasegawa2006}. In such cases, $D(\varepsilon)$ is proportional to $\sqrt{|\varepsilon-\varepsilon_{\rm F}|}$\cite{Dietl2008,Hasegawa2006}. 
When the semi-Dirac point exists at $\varepsilon_{\rm F}$ (referred to as semi-Dirac fermions), $C$ is proportional to $T^{1.5}$. 
On the other hand, when two Dirac points are close to each other, a crossover is expected changing from $C\propto T^{1.5}$ at high temperatures to $C\propto T^{2.0}$ at low temperatures. This is due to the distinct behavior of $D(\varepsilon)$ which is proportional to $\sqrt{|\varepsilon-\varepsilon_{\rm F}|}$ far from $\varepsilon_{\rm F}$ and to $|\varepsilon-\varepsilon_{\rm F}|$ near $\varepsilon_{\rm F}$. 
In this paper, we present numerical calculations demonstrating the occurrence of this crossover in $C$ when two Dirac points approach each other. We emphasize that this crossover serves as a signature of the topological phase transition.

\section{Calculation of electronic specific heat}

The internal energy per site ($U$) at temperature, $T$, can be calculated using the following equation:
\begin{equation}	
U(T)=\frac{1}{N}\sum_{i=1}^{m}\sum_{{\bf k}}
\varepsilon(i,{\bf k}) f(\varepsilon(i,{\bf k})), \label{U_1}
\end{equation}
where $\varepsilon(i,{\bf k})$ is an eigenvalue with band index $i$, $f[\varepsilon(i,{\bf k})]$ is the Fermi distribution function, 
$N=mN_k$ [$m=2$ for the honeycomb lattice and $m=4$ for $\alpha$-(BEDT-TTF)$_2$I$_3$], 
$N_k$ is the number of $\mathbf{k}$ points taken in the first Brillouin zone, and $k_{\rm B}$ is the Boltzmann constant. For simplicity, we neglect the effect of electron spin. 

The electronic specific heat, $C$, at the constant volume is obtained by differentiating the internal energy: 
\begin{eqnarray}
C=\frac{dU(T)}{dT}. \label{Ce}
\end{eqnarray}
We numerically calculate $C$ by performing numerical differentiation.

Using the density of states, $D(\varepsilon)$, Eq. (\ref{U_1}) can be 
expressed as
\begin{eqnarray}
U(T)=\frac{1}{N}\int_{-\infty}^{\infty}
(\varepsilon-\mu) D(\varepsilon)f(\varepsilon)d\varepsilon+\mu\nu,  
\label{U_3}
\end{eqnarray}
where $\mu$ is the chemical potential. 
When the $T$-dependence of $\mu$ is very small, it is known that $C\propto T$ for free electrons\cite{kittel}. In Section \ref{honey}, for two-dimensional systems we will show analytically and numerically demonstrate that Dirac fermions exhibit $C\propto T^{2.0}$ behavior, while semi-Dirac fermions show $C\propto T^{1.5}$ behavior. It is important to note that we are focusing solely on the {\it electronic} specific heat in this study, disregarding contribution from lattice vibrations ($\propto {T}^{3.0}$) that typically affect the specific heat in bulk metals\cite{kittel}.

For the purposes of this study, we do not consider electron-electron interactions. Although Coulomb interactions in graphene\cite{Vaf,She,Kot} have been proposed to introduce logarithmic corrections to thermodynamic properties\cite{Vaf}, these effects are expected to be small compared to the effects of the precursor of merging two Dirac points, which is the main focus of our investigation.

\begin{figure}[bt]
\begin{flushleft} \hspace{0.5cm}(a) \end{flushleft}\vspace{-0.5cm}
\includegraphics[width=0.38\textwidth]{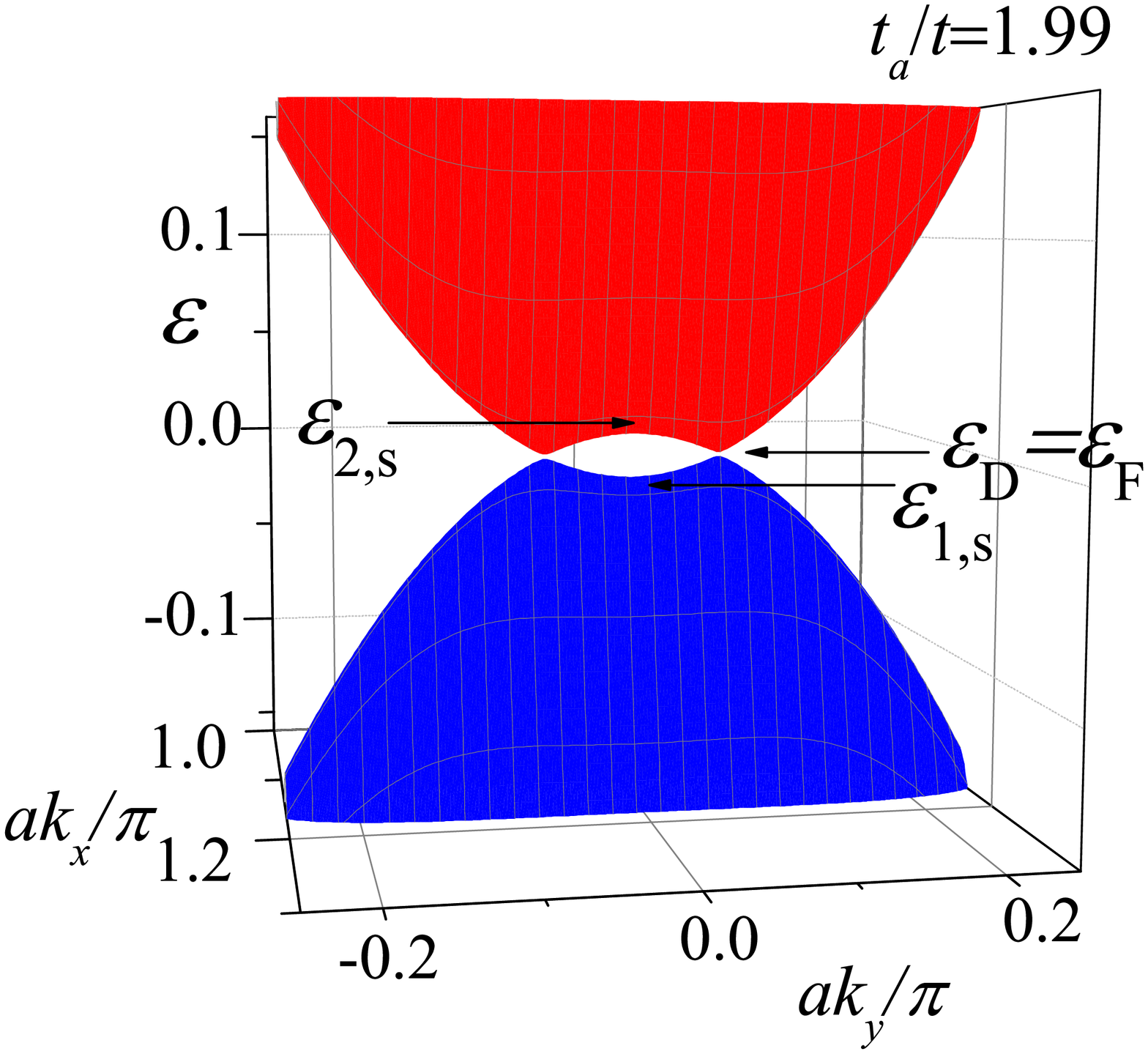}\vspace{-0.5cm}
\begin{flushleft} \hspace{0.5cm}(b) \end{flushleft}\vspace{-0.4cm}
\includegraphics[width=0.38\textwidth]{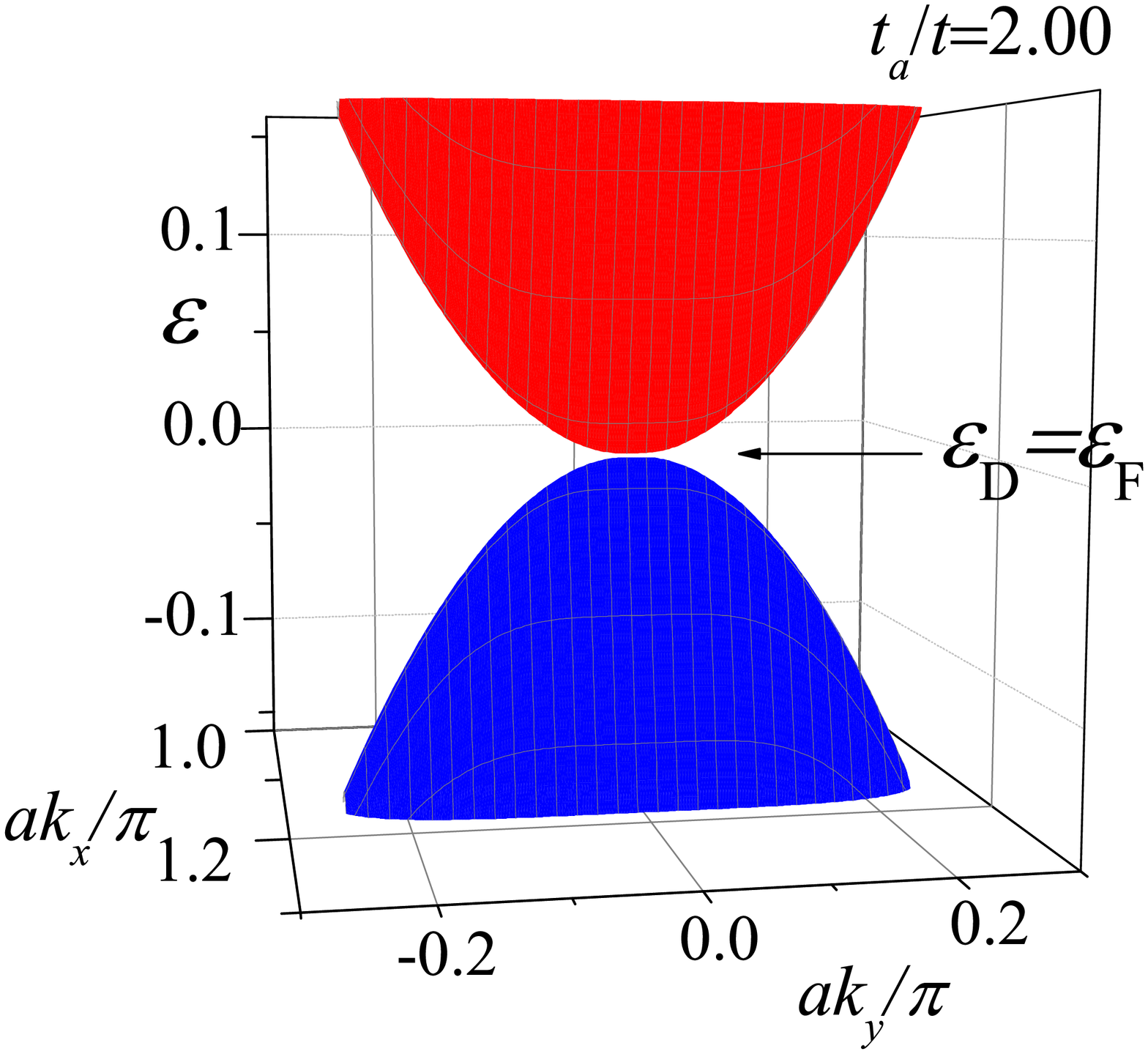}\vspace{-0.5cm}
\begin{flushleft} \hspace{0.5cm}(c) \end{flushleft}\vspace{-0.4cm}
\includegraphics[width=0.38\textwidth]{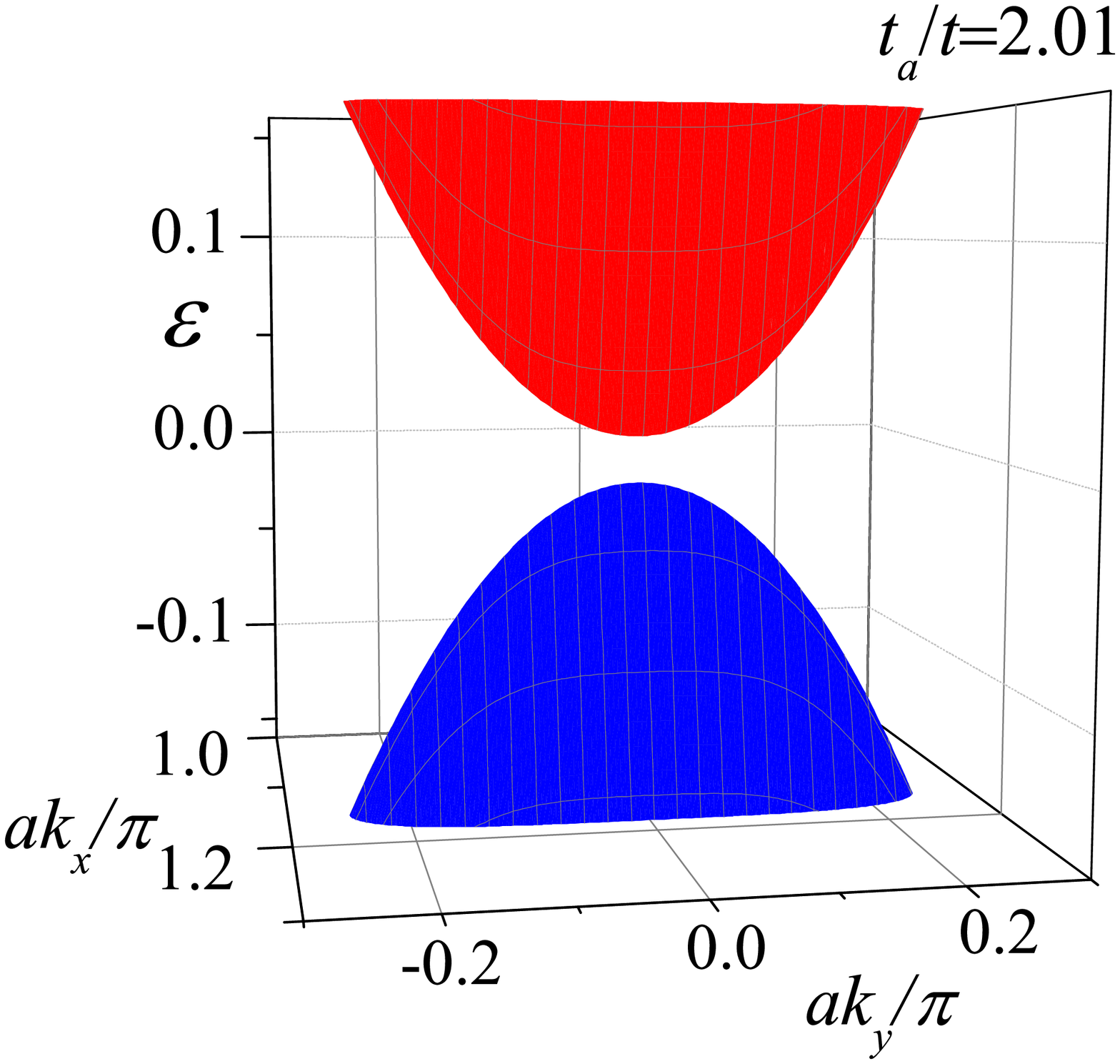}\vspace{-0.0cm}
\caption{
 (Color online) 
Upper and lower bands for different values of $t_a$ in the honeycomb lattice. (a) $t_a/t=1.99$, (b) $t_a/t=2.00$, and (c) $t_a/t=2.01$. At $t_a/t\leq 2.00$, the Fermi energy, $\varepsilon_{\rm F}$, is zero, coinciding with $\varepsilon_{\rm D}$. 
}
\label{fig_h_d}
\end{figure}

\begin{figure}[bt]
\vspace{-0.2cm}
\includegraphics[width=0.51\textwidth]{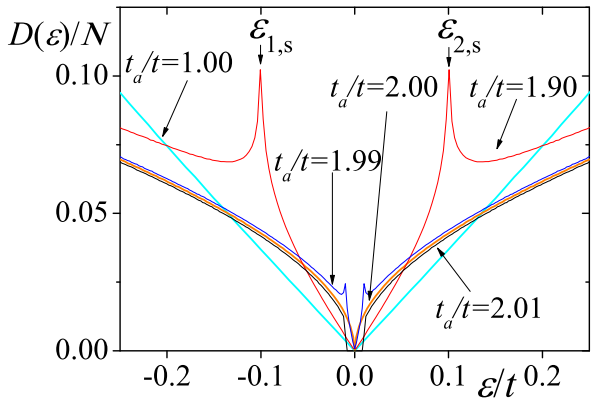}\vspace{-0.2cm}
\caption{
 (Color online) 
Densities of the states at different $t_a$ values, where $t_a/t=1.00$ (a sky blue line), 1.90 (a red line), 1.99 (a blue line), 2.00 (an orange line), and $2.01$ (a black line). 
At $t_a/t=1.00$ (Dirac fermions), $D(\varepsilon) \propto |\varepsilon|$ and at $t_a/t=2.00$ (semi-Dirac fermions), $D(\varepsilon) \propto \sqrt{|\varepsilon|}$, respectively. At $t_a/t=2.01$ (a normal insulator), $D(\varepsilon)$ near $\varepsilon_{\rm F}$ becomes zero due to the energy gap. 
}
\label{fig_h_e}
\end{figure}

\section{Energy band in the honeycomb lattice}
\label{EB}

We consider the honeycomb lattice with only nearest neighbor transfer integrals ($t_a$, $t_b$, and $t_c$) and half-filling. 
The energy dispersion relations are given by\cite{wallace} 
\begin{eqnarray}
\varepsilon(\pm,{\bf k})&=&\pm \bigg[t_{a}^{2}+t_{b}^{2}+t_{c}^{2}+ 2t_a t_b\cos( \frac{\sqrt{3}}{2}a k_x - \frac{1}{2}ak_y )\nonumber \\
&+& 2t_a t_c\cos( \frac{\sqrt{3}}{2}a k_x + \frac{1}{2}ak_y ) + 2t_b t_c\cos( ak_y )\bigg]^{\frac{1}{2}},\label{g1}
\end{eqnarray} 
where $\pm$ signs in $\varepsilon(\pm,{\bf k})$ correspond to upper and lower bands, respectively, and $a$ represents the lattice constant. We assume $t_b=t_c=t$ and $t_a\geq t$ in Eq. (\ref{g1}), where $t>0$. 
The Fermi energy, $\varepsilon_{\rm F}$, is equal to the energy at the Dirac point ($\varepsilon_{\rm D}$) because of the half filling. Due to the particle-hole symmetry exists in Eq. (\ref{g1}), $\mu$ remain constant as a function of $T$ (i.e., $\mu=\varepsilon_{\rm F}=\varepsilon_{\rm D}=0$).

The position of two Dirac points is close each other at $t_a/t=1.99$, as shown in Fig. \ref{fig_h_d} (a). They merge at $t_a/t=2.00$, and an energy gap opens at $t_a/t>2.00$, as depicted in Figs. \ref{fig_h_d} (b) and (c), respectively. The spacing between two Dirac points ($\Delta k$) decreases as $t_a$ increases. Fig. \ref{fig_h_d} (a) also shows a saddle point in the upper and lower bands at the midpoint between two Dirac points. 
The energies ($\varepsilon_{2,{\rm s}}$ and $\varepsilon_{1,{\rm s}}$) at the saddle point for the upper and lower bands, respectively, are given by 
\begin{eqnarray} 
\varepsilon_{1,{\rm s}}/t=-\varepsilon_{2,{\rm s}}/t=t_a/t-2.00. \label{es}
\end{eqnarray}

For Dirac fermions ($t_a/t=1.00$), the energies near $\varepsilon_{\rm F}$ can be described as\cite{wallace} 
\begin{eqnarray}
\varepsilon(\pm,{\bf k}) =\pm \hbar v_{\rm{F}} \sqrt{ k_x^2 +k_y^2}, \label{E_D}
\end{eqnarray}
where $\hbar=h/2\pi$, $h$ is the Planck constant, and 
$v_{\rm{F}}=\sqrt{3} a t/(2\hbar)$. The density of states, $D(\varepsilon)$, for Eq. (\ref{E_D}) is given by\cite{wallace}
\begin{eqnarray}
D(\varepsilon)=\frac{2{N}}{\sqrt{3}\pi t^{2} } |\varepsilon|.\label{DD}
\end{eqnarray}
For semi-Dirac fermions ($t_a/t=2.00$), the energies near $\varepsilon_{\rm F}$ can be written as\cite{Dietl2008,Hasegawa2006}
 \begin{eqnarray}
 \varepsilon(\pm,{\bf k}) =\pm \hbar v_{\rm{F}} \sqrt{k_x^2 +\alpha^2k_y^4},
\end{eqnarray}
where 
$\alpha =a/(4\sqrt{3})$. The dispersion is linear in $\pm k_x$-directions and quadratic in $\pm k_y$-directions. 
The density of states is given by\cite{Hasegawa2006}
\begin{eqnarray}
D(\varepsilon)= \frac{N}{4(\pi t)^{\frac{3}{2}}}  \frac{\Gamma(\frac{1}{4})}{\Gamma (\frac{3}{4})}\sqrt{|\varepsilon|}. \label{d_semi}
\end{eqnarray}
We numerically calculate $D(\varepsilon)$ using Eq. (\ref{g1}), as shown in Fig. \ref{fig_h_e}. At $t_a/t=1.90$ and $1.99$, the density of states, $D(\varepsilon)$, is proportional to $|\varepsilon|$ when $\varepsilon$ is near $\varepsilon_{\rm F}$, and it approaches $D(\varepsilon)\propto\sqrt{|\varepsilon|}$ above $\varepsilon_{2,{\rm s}}$ and below $\varepsilon_{1,{\rm s}}$. The two peaks in $D(\varepsilon)$ are due to the energies at the saddle points.

\begin{figure}[bt]
\vspace{-0.2cm}
\includegraphics[width=0.51\textwidth]{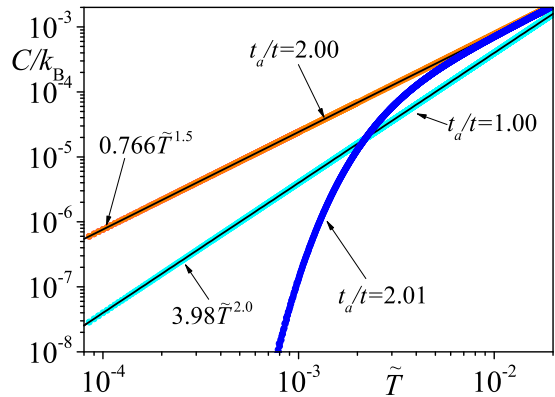}\vspace{-0.2cm}
\caption{
 (Color online) 
The electronic specific heats as a function of ${\widetilde T}$ from numerical calculations. The electronic specific heats at $t_a/t=1.00$, at $t_a/t=2.00$, and at $t_a/t=2.01$ are represented by sky blue circles, orange circles, and blue circles. The black solid lines corresponds to the theoretical predictions of Eq. (\ref{C_D}) and Eq. (\ref{C_S}), respectively. 
}
\label{fig_12}
\end{figure}

\begin{figure}[bt]
\begin{flushleft} 
\hspace{0.5cm}(a) \end{flushleft}\vspace{-0.2cm}
\includegraphics[width=0.52\textwidth]{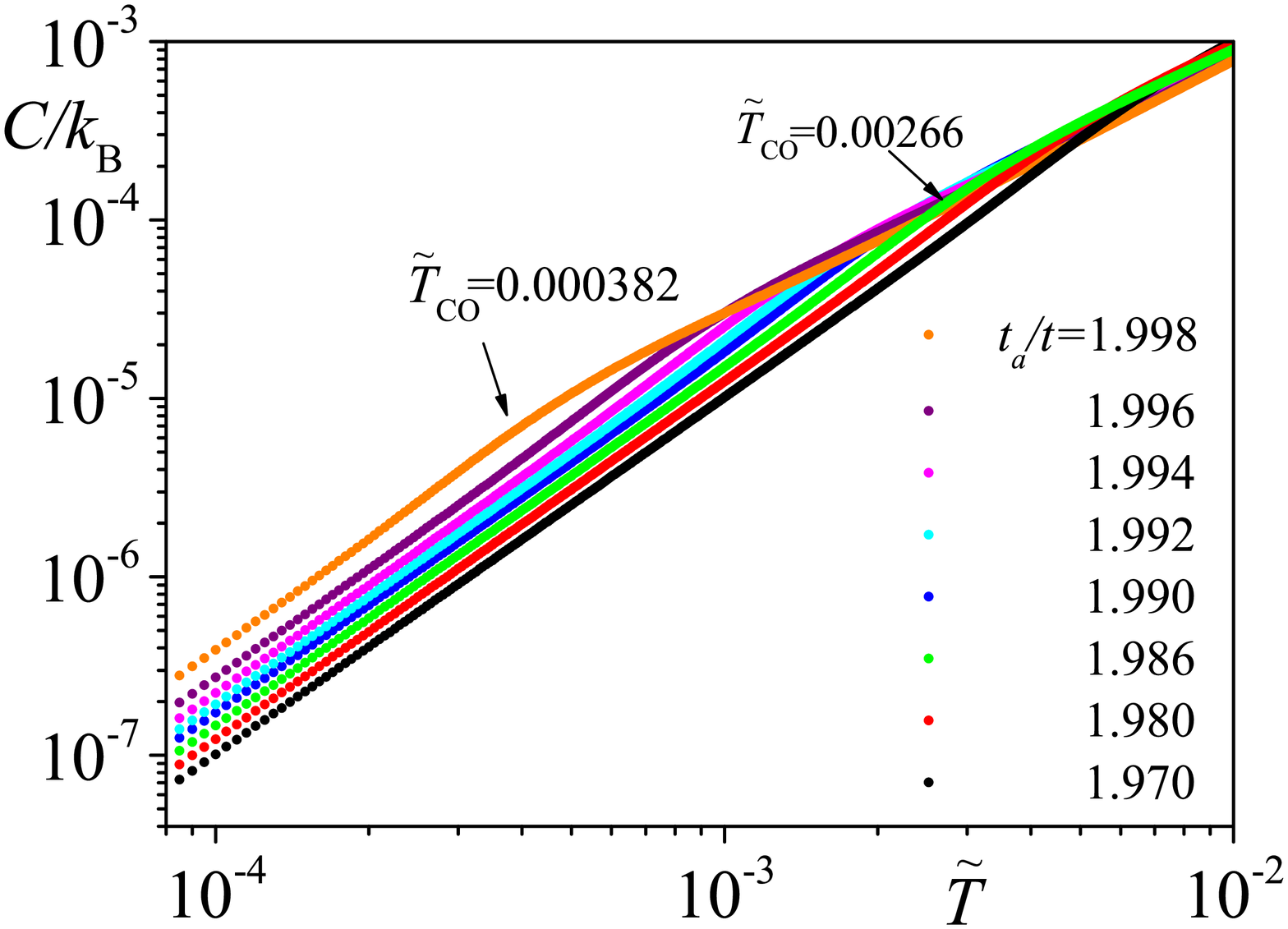}\vspace{-0.5cm}
\begin{flushleft} \hspace{0.5cm}(b) \end{flushleft}\vspace{-0.3cm}
\includegraphics[width=0.51\textwidth]{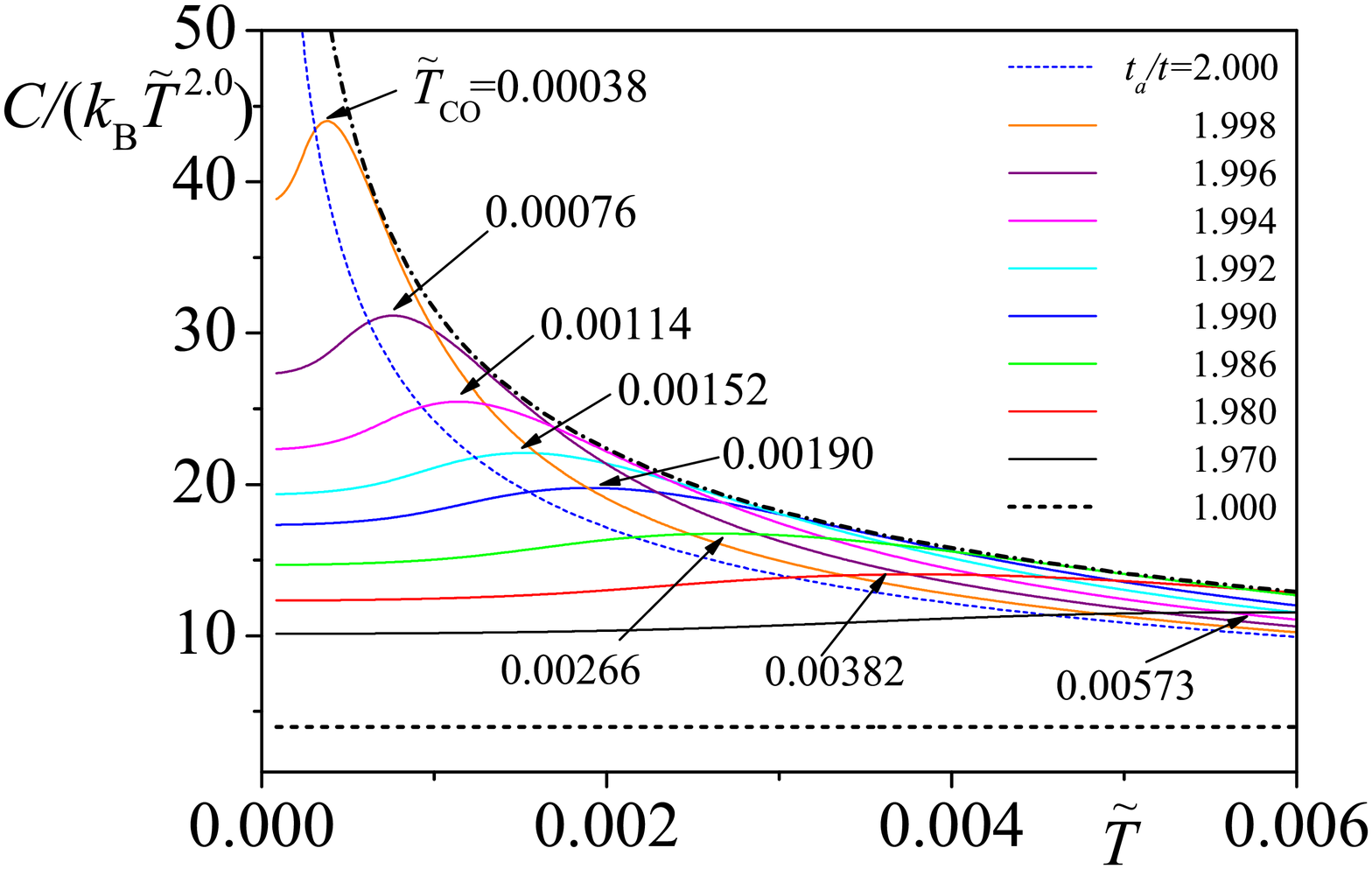}\vspace{-0.0cm}
\caption{
 (Color online)
(a) Electronic specific heats at different $t_a$ values as a function of ${\widetilde T}$, where ${\widetilde T}_{\rm co}$ are for $t_a/t=1.998$ and 1.986 in (b). (b) $C/(k_{\rm B}{\widetilde T}^{2.0})$ as a function of ${\widetilde T}$, where a dot-dashed line represents $\propto 1/\sqrt{{\widetilde T}}$. 
}
\label{fig_13}
\end{figure}

\begin{figure}[bt]
\begin{flushleft} 
\end{flushleft}\vspace{-0.0cm}
\includegraphics[width=0.5\textwidth]{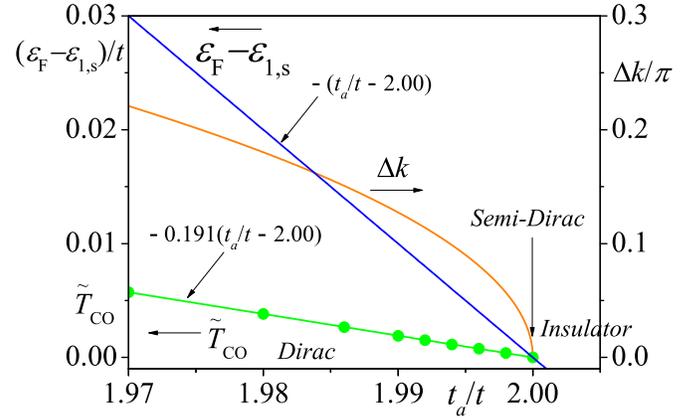}\vspace{-0.0cm}
\caption{ (Color online) 
$\Delta k$, $\varepsilon_{1,{\rm s}}$, and ${\widetilde T}_{\rm CO}$ as a function of $t_a$ which are shown by an orange line, a blue line, and green circles, respectively. A green line is $-0.191(t_a/t-2.00)$. 
}
\label{fig13_2}
\end{figure}

\section{Electronic specific heat in the honeycomb lattice}
\label{honey}

In the honeycomb lattice, the electronic specific heat, $C$, can be calculated for both Dirac fermions and semi-Dirac fermions. 
For Dirac fermions, using the density of states $D(\varepsilon)$ given in Eq. (\ref{DD}), we can analytically derive the electronic specific heat as follows: 
\begin{eqnarray}
C=\frac{6\sqrt{3}}{\pi}\zeta(3)k_{\rm B}\widetilde{T}^{2.0}\simeq3.98 k_{\rm B}\widetilde{T}^{2.0}, \label{C_D}
\end{eqnarray}
where $\widetilde{T}=k_{\rm B}T/t$ is the dimensionless temperature. 
For example, in the case of graphene, $t$ can be estimated to be around 2.97 eV from first-principles calculations\cite{Reich}. Thus, $\widetilde{T}=0.0001$ corresponds to approximately 3.45 K.

For semi-Dirac fermions, using $D(\varepsilon)$ given in Eq. (\ref{d_semi}),  we obtain the following expression for the electronic specific heat, $C$, as 
\begin{eqnarray}
C=\frac{15(4-\sqrt{2})\zeta(\frac{5}{2})\Gamma(\frac{1}{4})}
{64\pi\Gamma(\frac{3}{4})}k_{\rm B}\widetilde{T}^{1.5}\simeq0.766 k_{\rm B}\widetilde{T}^{1.5}. \label{C_S}
\end{eqnarray}
We compare these analytical expressions for $C$ with numerically calculated $C$, as shown in Fig. \ref{fig_12}. It can be seen that the numerical results are well-fitted by Eq. (\ref{C_D}) for Dirac fermions and Eq. (\ref{C_S}) for semi-Dirac fermions, respectively. In the case where $t_a/t=2.01$, $C$ is suppressed at low temperatures due to the system becoming insulating.


Next, we consider the cases when two Dirac points exist close each other. 
We perform numerical calculations of $C$ at $1.970\leq t_a/t\leq 1.998$, as shown in Fig. \ref{fig_13} (a). The results exhibit changes in the power-law dependence on temperature. To clarify the deviations from Dirac fermions, we plot $C/{\widetilde T}^{2.0}$ in Figs. \ref{fig_13} (b). It can be observed that $C/{\widetilde T}^{2.0}$ remains almost constant at low temperatures and becomes proportional to $1/\sqrt{{\widetilde T}}$ at high temperatures. 
Thus, $C$ is nearly proportional to ${\widetilde T}^{2.0}$ at low temperatures and to ${\widetilde T}^{1.5}$ at high temperatures, respectively. There is a crossover in the ${\widetilde T}$-dependence of $C$, which occurs due to the $\varepsilon$-dependence of $D(\varepsilon)$. The contribution to $C$ at low temperatures arises from $D(\varepsilon)$ near $\varepsilon_{\rm F}$, while  at high temperatures,  it comes from the energy range above $\varepsilon_{2,{\rm s}}$ and below $\varepsilon_{1,{\rm s}}$. The temperature at which this crossover occurs decreases as $t_a$ increases, approaching $t_a/t=2.000$. It is not possible to determine an exact crossover temperature, because $C$ varies continuously from $C\propto {\widetilde T}^{2.0}$ to $C\propto {\widetilde T}^{1.5}$ as a function of ${\widetilde T}$. 
We observe the existence of maxima in $C/{\widetilde T}^{2.0}$, except for the cases of $t_a/t=1.000$ and 2.000, as shown in Fig. \ref{fig_13}(b). These maxima arise from the logarithmic divergence of $D(\varepsilon)$ due to the energy at the saddle point, and the crossover occurs near these maxima. Therefore, in this paper, we employ the temperature at which $C/{\widetilde T}^{2.0}$ is maximized as a rough estimation of the crossover temperature (${{\widetilde T}}_{\rm co}$).

The crossover temperature, ${{\widetilde T}}_{\rm co}$, the difference in wave vectors $\Delta k$, and the energy difference $\varepsilon_{\rm F}-\varepsilon_{1,{\rm s}}$ are shown as functions of $t_a/t$ in Fig. \ref{fig13_2}. We find that ${{\widetilde T}}_{\rm co}$ can be fitted by the expression:
\begin{eqnarray}
{{\widetilde T}}_{\rm co}=-0.191(t_a/t-2.00), \label{Tco}
\end{eqnarray}
which is depicted by the green line. As $t_a$ increases, 
${{\widetilde T}}_{\rm co}$, $\varepsilon_{\rm F}-\varepsilon_{1,{\rm s}}$, and $\Delta k$ decrease. The magnitude of ${{\widetilde T}}_{\rm co}$ is on the order of the potential barrier between the two Dirac points. When the barrier is zero (i.e., in the case of semi-Dirac fermions), ${T}_{\rm co}$ becomes zero.

\newpage

\section{Energy band of 
$\alpha$-(BEDT-TTF)$_2$I$_3$ under the uniaxial pressure}
\label{EB_I3}

In the study of the energy band of $\alpha$-(BEDT-TTF)$_2$I$_3$ under uniaxial pressure, we consider a tight-binding model with a rectangular lattice for the highest occupied molecular orbits of the BEDT-TTF molecule. The electron filling ($\nu$) of this material is 3/4. 
We neglect the interlayer hopping, $t_{\perp}$, because of its small magnitude. In two-dimensional conductive plane, we incorporate the transfer integrals between neighboring sites ($t_{\mathrm{a1}}$, $t_{\mathrm{a2}}$, $t_{\mathrm{a3}}$, $t_{\mathrm{b1}}$, $t_{\mathrm{b2}}$, $t_{\mathrm{b3}}$, and $t_{\mathrm{b4}}$). 
The Hamiltonian matrix has been explained in Appendix C in ref \onlinecite{KH2017}. The effect of the uniaxial pressure ($P$) along the $y$-axis is taken into account by modifying the values of the transfer integrals using interpolation\cite{Katayama2004,Katayama2006,Suzumura2013} based on the extended H\"uckel method\cite{Mori1984,Kondo2005}, with the units of transfer integrals and pressure being in eV and kbar, respectively.  

\begin{figure}[bt]
\begin{flushleft} 
\hspace{0.5cm}(a) 
\end{flushleft}\vspace{-0.5cm}
\includegraphics[width=0.39\textwidth]{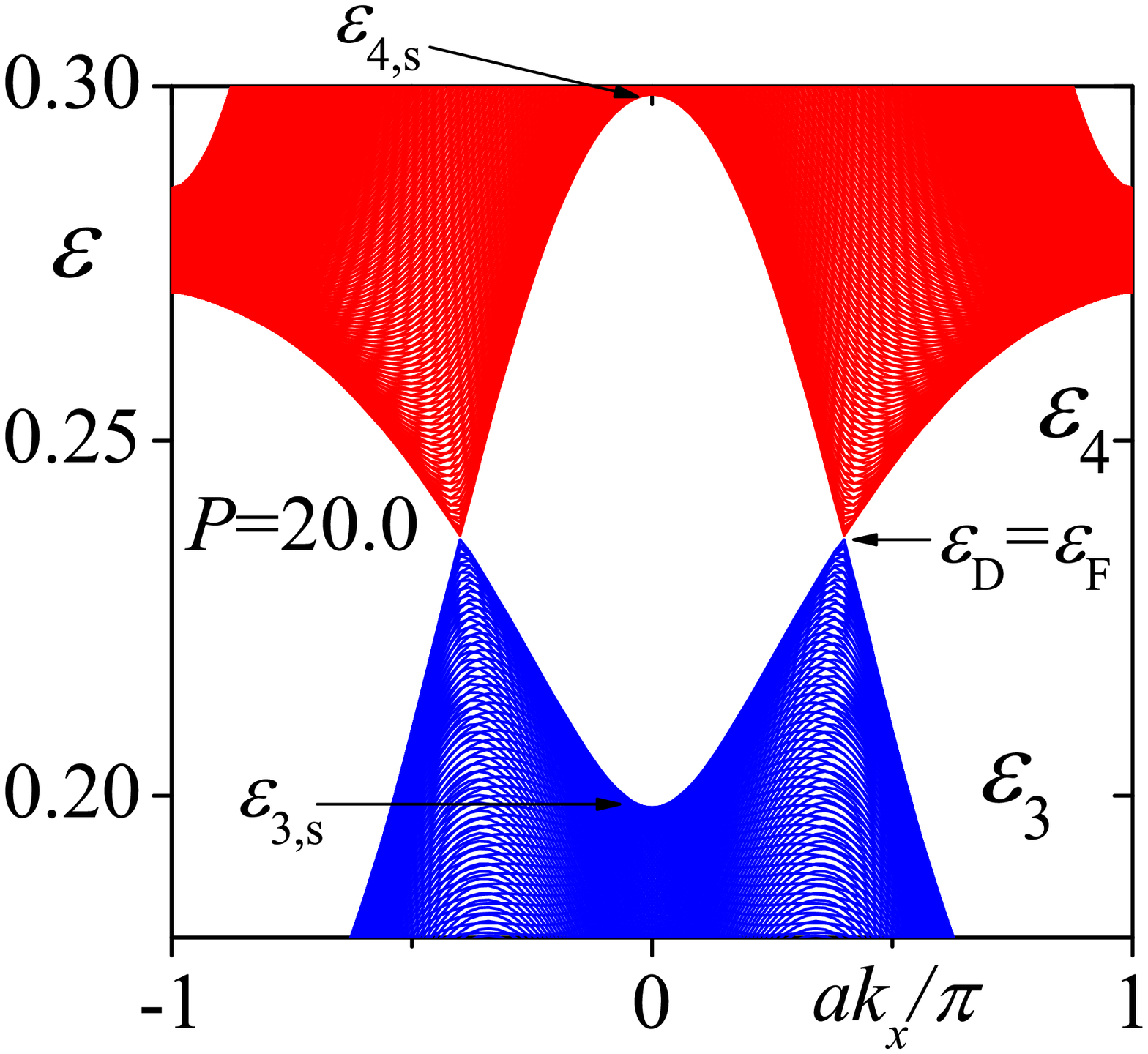}\vspace{-0.5cm}
\begin{flushleft} \hspace{0.5cm}(b) \end{flushleft}\vspace{-0.5cm}
\includegraphics[width=0.39\textwidth]{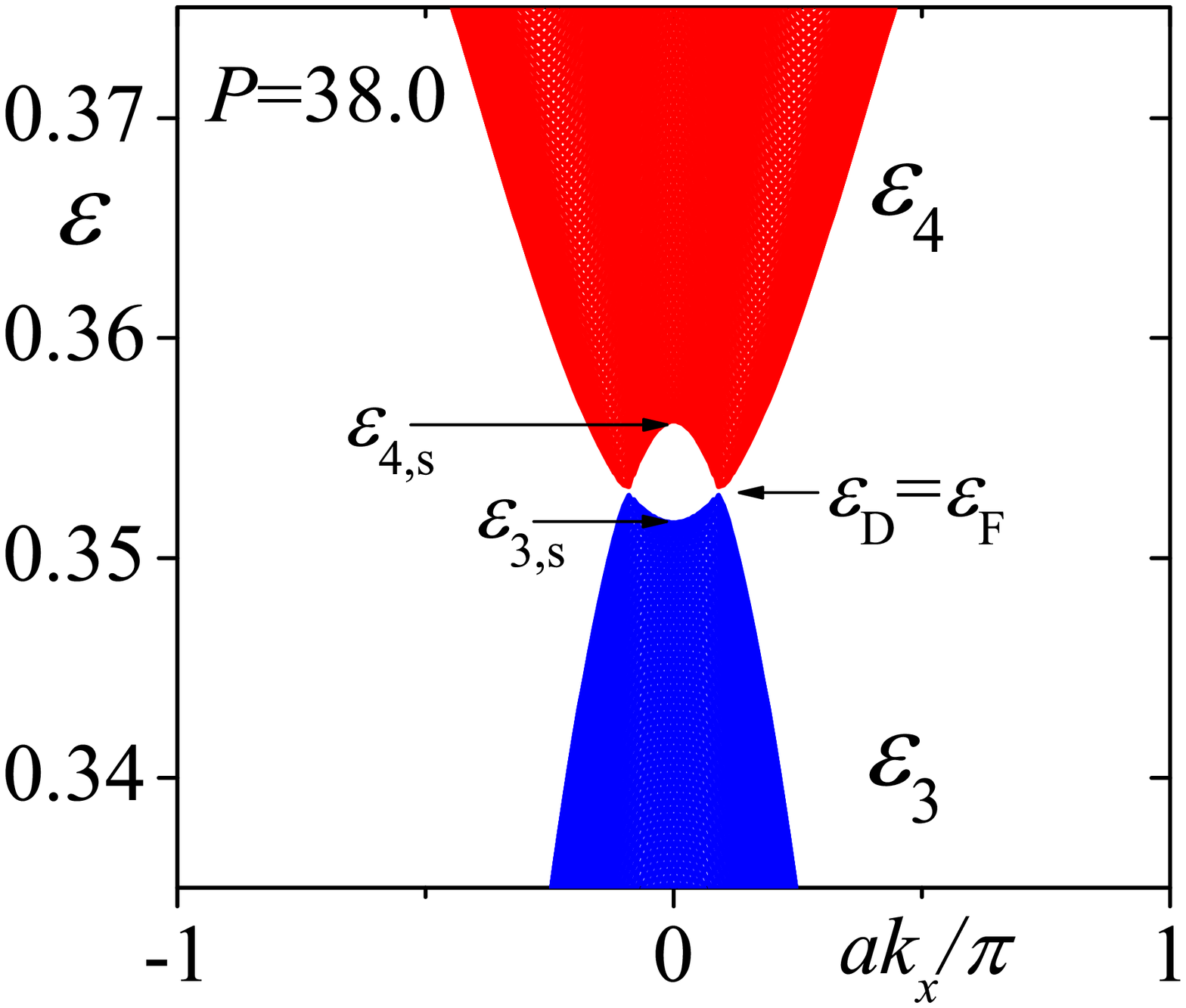}\vspace{-0.5cm}
\begin{flushleft} \hspace{0.5cm}(c) \end{flushleft}\vspace{-0.5cm}
\includegraphics[width=0.39\textwidth]{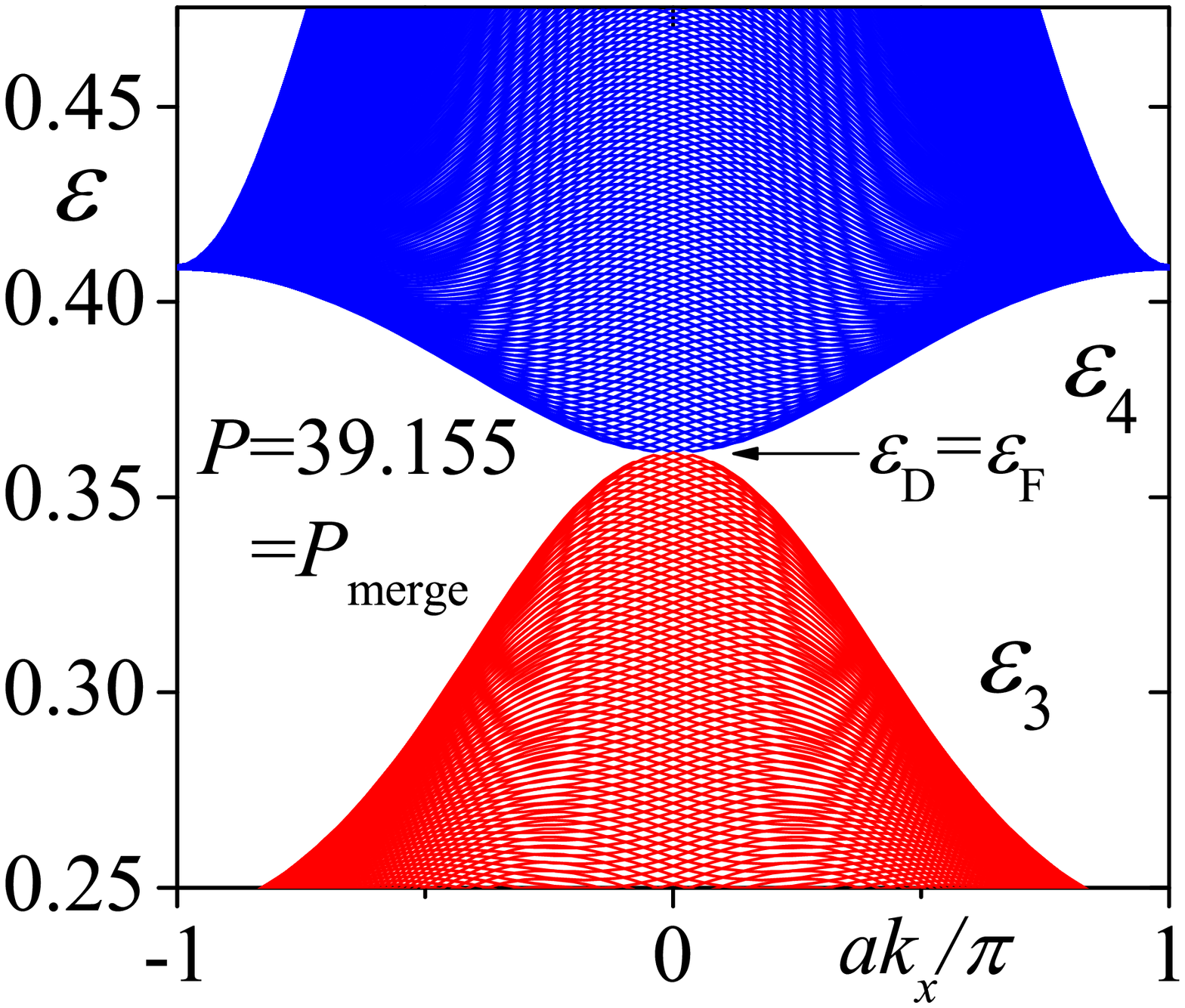}\vspace{-0.5cm}
\begin{flushleft} \hspace{0.5cm}(d) \end{flushleft}\vspace{-0.5cm}
\includegraphics[width=0.39\textwidth]{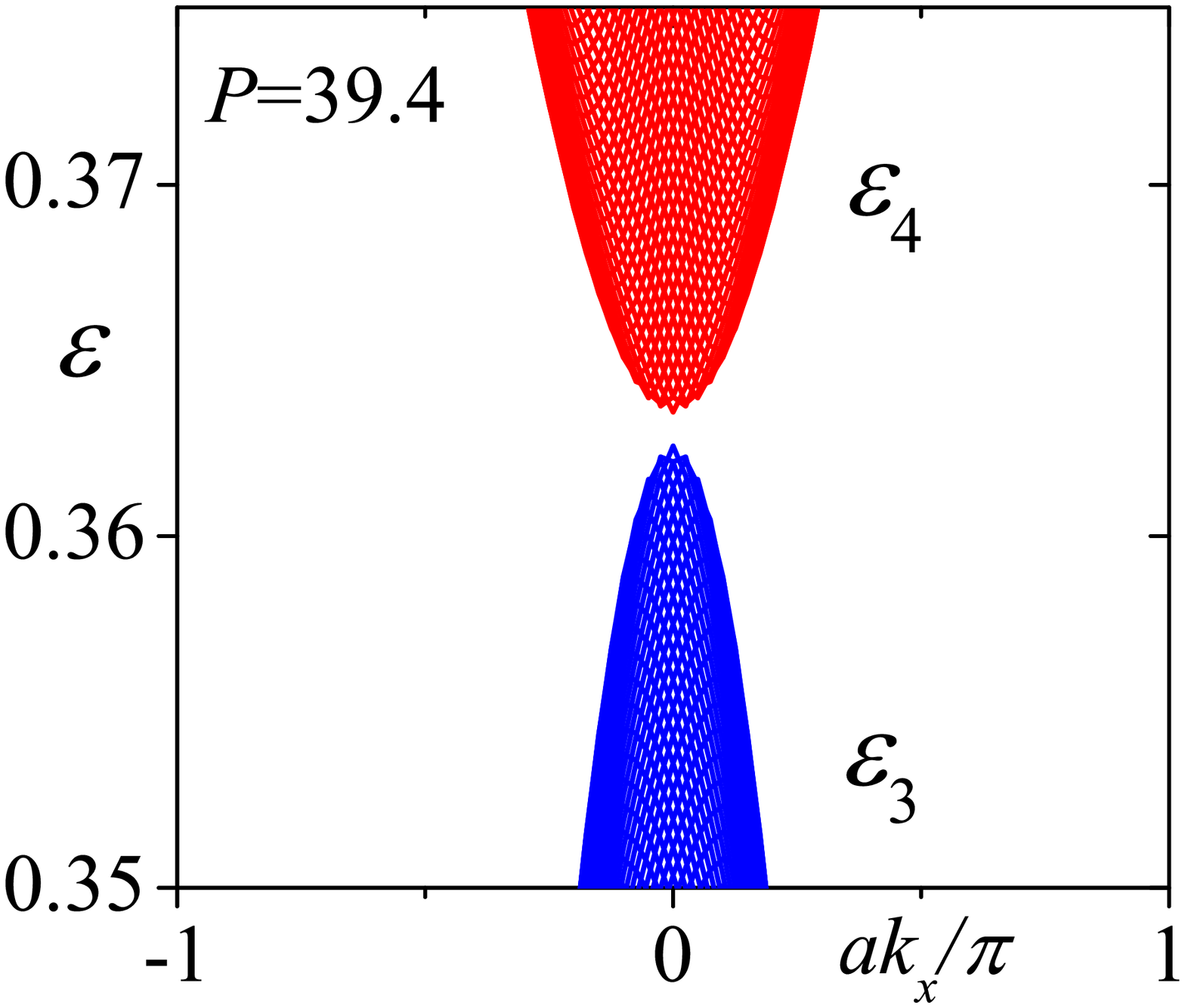}\vspace{0.1cm}
\caption{ (Color online) 
The third and fourth energy bands ($\varepsilon_{\rm 3}$ and 
$\varepsilon_{\rm 4}$) from the bottom at $P=20.0$ (a), $38.0$ (b), $39.155$ (c), and 39.4 (d), where the view point is $(k_x, k_y)=(0,-\infty)$ and $a$ is the lattice constant along $x$-direction. The energies at the saddle point in the third and fourth bands are denoted by $\varepsilon_{\rm 3, s}$ and $\varepsilon_{\rm 4, s}$, respectively.
}
\label{fig2_b}
\end{figure}

In this model, two Dirac cones are
overtilted (type-II Dirac or Weyl semimetal) at $0\leq P< 2.3$\cite{KH2017}.
At $P>2.3$ the Dirac cones are tilted but not overtilted (type-I Dirac or Weyl semimetal), and the energy at the Y point $(0, \pi)$ in the third band from the bottom is higher than the energies at the Dirac points, resulting in an overlap between the third and the fourth bands. This overlapping occurs within $0\leq P< 3.0$. 
At $P> 3.0$, these bands no longer overlap, and the system supports the tilted Dirac cones of type I\cite{Katayama2006}. Fig. \ref{fig2_b} (a) illustrates this behavior for a specific value, $P=20.0$. The Fermi energy, $\varepsilon_{\rm F}$, is located at $\varepsilon_{\rm D}$. Furthermore, as $P$ increases, the spacing between two Dirac points, $\Delta k$, becomes smaller. This can be observed in Fig. \ref{fig2_b} (b) for $P=38.0$. Upon reaching a critical pressure $P_{\rm merge}=39.155$, two Dirac points merge, resulting in a quadratic dispersion along the $\pm k_x$-directions, as shown in Fig. \ref{fig2_b} (c). 
At $P>P_{\rm merge}$, a finite gap emerges in the energy band, as shown in Fig. \ref{fig2_b} (d) for $P=39.4$.

\begin{figure}[bt]
\begin{center}
\vspace{-0.2cm}
\includegraphics[width=0.51\textwidth]{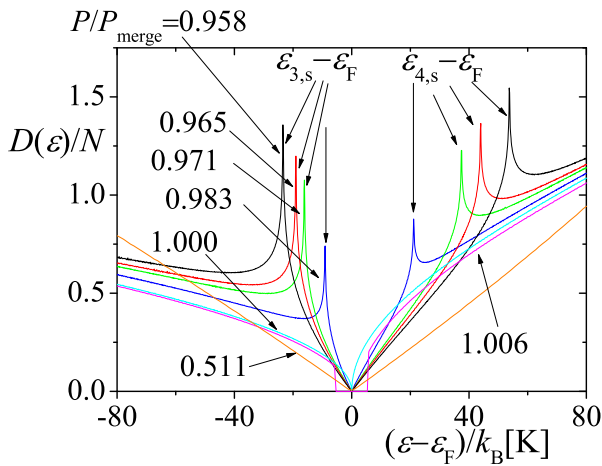}\vspace{-0.0cm}
\end{center}
\caption{
 (Color online) 
The densities of states at different $P$ values as a function of $(\varepsilon-\varepsilon_{\rm F})/k_{\rm B}$, respectively. The densities of states are asymmetric with respect to $\varepsilon _{\rm F}$ because of no particle-hole symmetry between third and fourth bands. There are two peaks due to the energy at the saddle points at $0.958\leq P/P_{\rm merge}\leq 0.983$.
}
\label{fig_7}
\end{figure}

For semi-Dirac fermions ($P=P_{\rm merge}$) and Dirac fermions ($P=$20.0, i.e., $P/P_{\rm merge}=0.511$), the densities of states follows $D(\varepsilon)\propto \sqrt{|\varepsilon-\varepsilon _{\rm F}|}$ and $D(\varepsilon)\propto|\varepsilon-\varepsilon_{\rm F}|$, respectively, as shown in Fig. \ref{fig_7}. 
When two Dirac points are close each other 
($P=37.5, 37.8, 38.0$, and 38.5, i.e., $P/P_{\rm merge}=0.958, 0.965, 0.971$, and 0.983), the density of states is proportional to $|\varepsilon-\varepsilon_{\rm F}|$ when $\varepsilon$ is close to $\varepsilon_{\rm F}$ and 
it approaches $D(\varepsilon)\propto\sqrt{|\varepsilon-\varepsilon_{\rm F}|}$ when $\varepsilon$ exceeds the peaks at $\varepsilon_{3,s}$ or $\varepsilon_{4,s}$ at these pressures, as shown in Fig. \ref{fig_7}. 
For insulating state ($P=39.4$, i.e., $P/P_{\rm merge}=1.006$), $D(\varepsilon)$ near $\varepsilon _{\rm F}$ is zero, as shown in Fig. \ref{fig_7}.

\section{Electronic specific heat of 
$\alpha$-(BEDT-TTF)$_2$I$_3$ under the uniaxial pressure}
\label{h=0}

\begin{figure}[bt]
\vspace{0.5cm}
\includegraphics[width=0.5\textwidth]{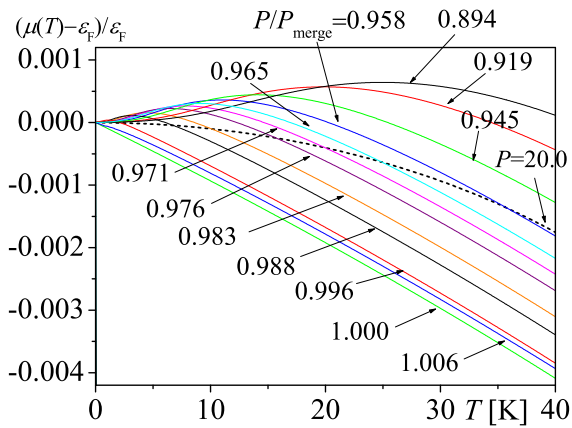}\vspace{-0.0cm}
\caption{
 (Color online) 
The chemical potential at different $P$ values as a function of $T$. 
}
\label{fig_8}
\end{figure}

\begin{figure}[bt]
\vspace{-0.2cm}
\includegraphics[width=0.51\textwidth]{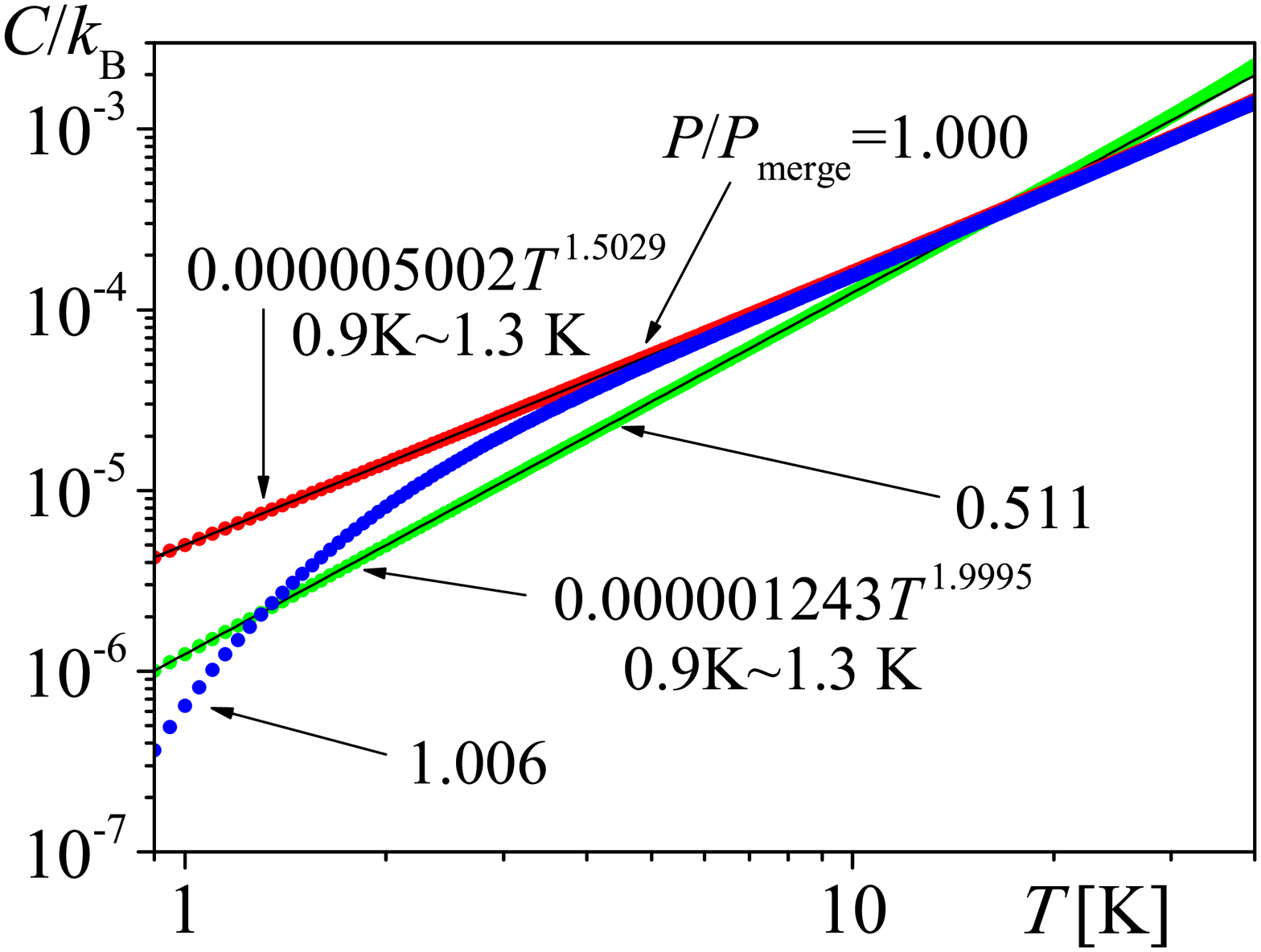}\vspace{-0.0cm}
\caption{
 (Color online) 
Electronic specific heats as a function of $T$ at $P/P_{\rm merge}=0.511$ (green circles),  $P/P_{\rm merge}=1.000$ (red circles), and $P/P_{\rm merge}=1.006$ (blue circles), respectively. 
Black lines are fitting lines of $0.000005002T^{1.5029}$ and $0.000001243T^{1.9995}$, respectively. 
}
\label{fig_10}
\end{figure}

\begin{figure}[bt]
\begin{flushleft} \hspace{0.5cm}(a) \end{flushleft}\vspace{-0.5cm}
\includegraphics[width=0.5\textwidth]{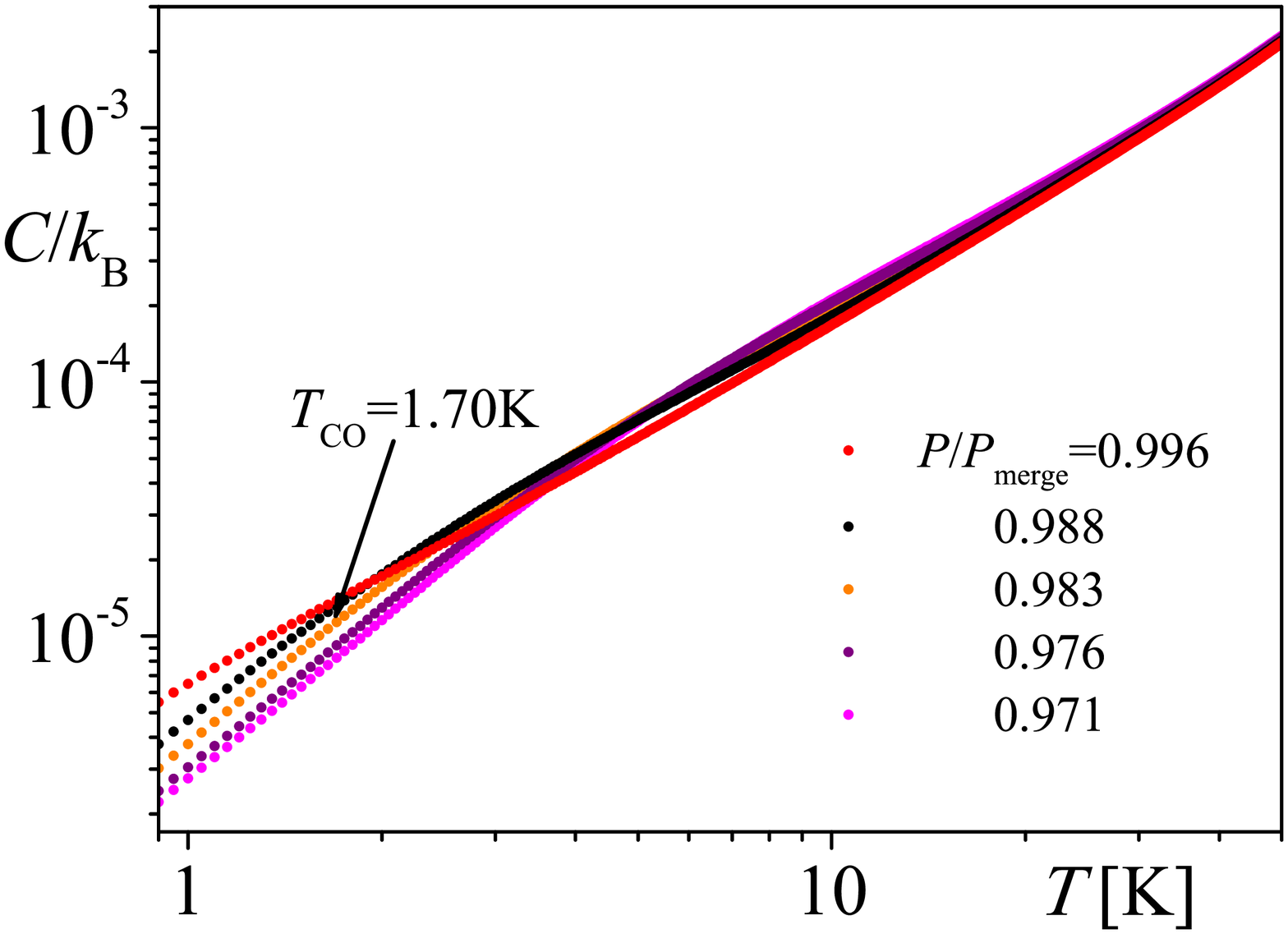}\vspace{-0.0cm}
\begin{flushleft} \hspace{0.5cm}(b) \end{flushleft}\vspace{-0.5cm}
\includegraphics[width=0.5\textwidth]{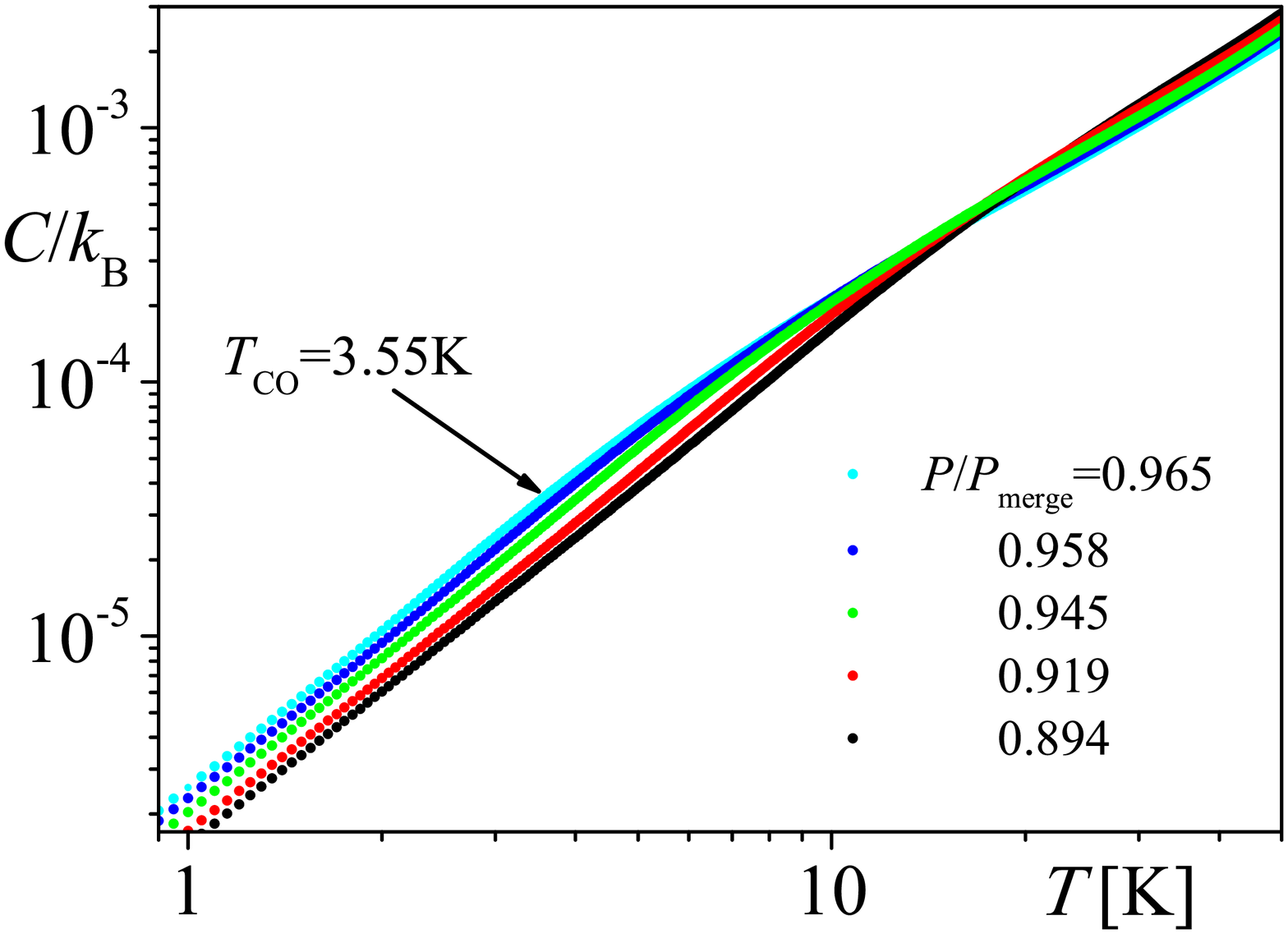}\vspace{-0.0cm}
\caption{
 (Color online) (a) and (b) Electronic specific heats at different $P$ values as a function of $T$, where $T_{\rm co}$ are for $P/P_{\rm merge}=0.983$ and 0.965 in Fig. \ref{fig_10_4}. 
}
\label{fig_10_3}
\end{figure}

\begin{figure}[bt]
\vspace{-0.2cm}
\includegraphics[width=0.5\textwidth]{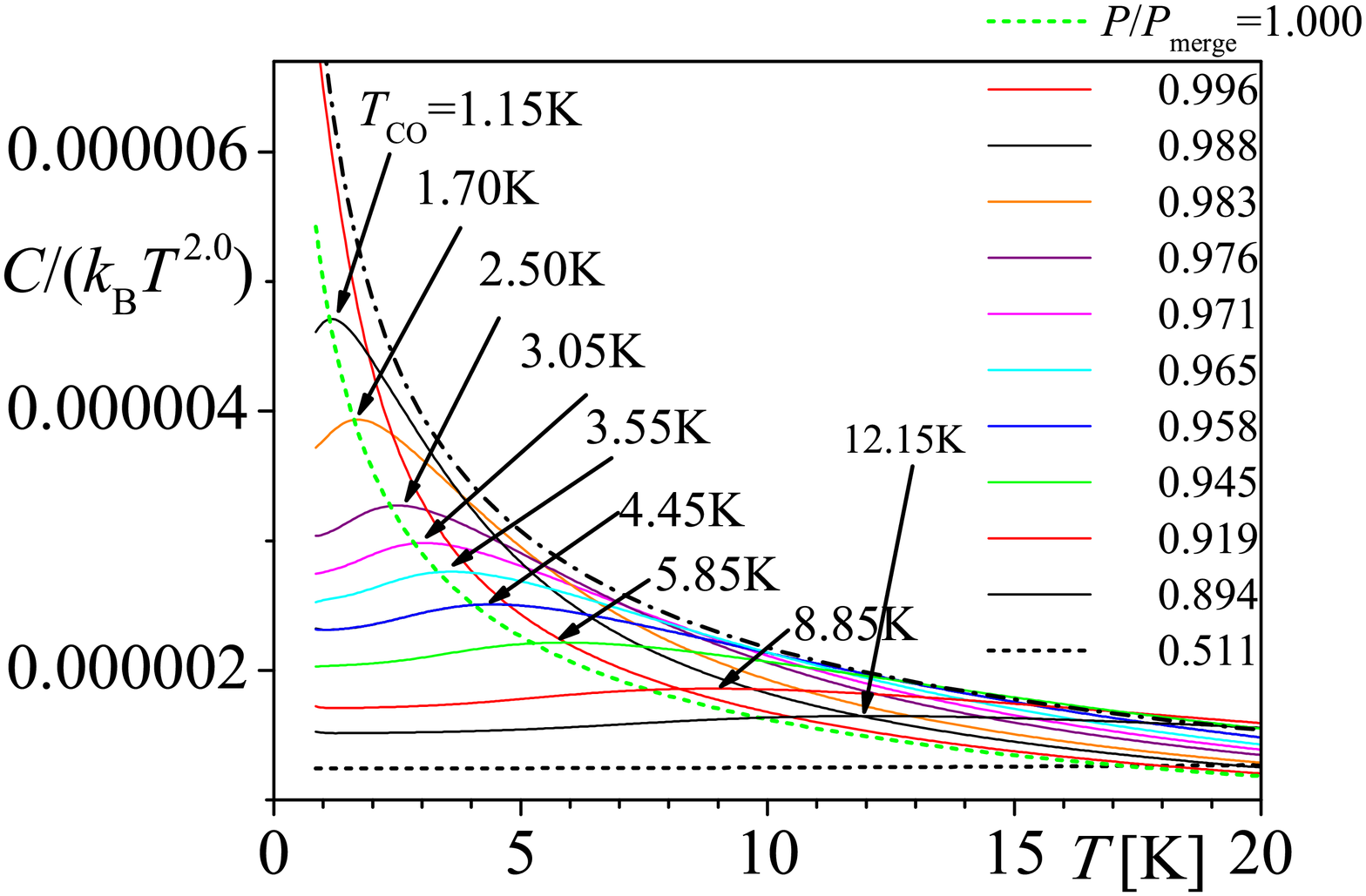}\vspace{0.0cm}
\caption{
 (Color online) $C/(k_{\rm B}T^{2.0})$ at different $P$ values as a function of $T$, where a dot-dashed line represents $\propto 1/\sqrt{T}$.}
\label{fig_10_4}
\end{figure}

\begin{figure}[bt]
\vspace{-0.2cm}
\includegraphics[width=0.5\textwidth]{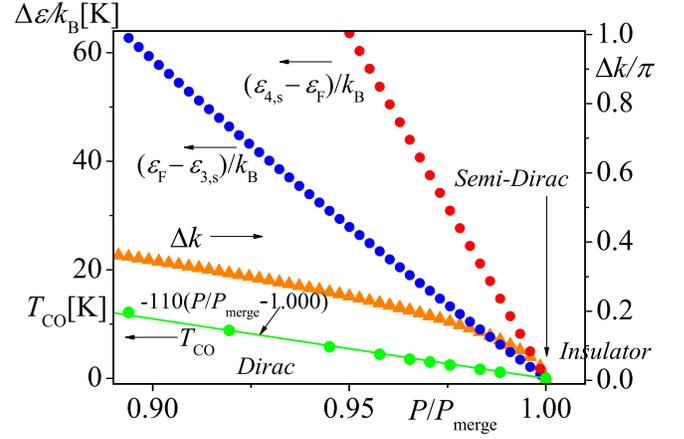}\vspace{0.0cm}
\caption{
 (Color online) 
Green circles, orange triangles, red circles and blue circles are $T_{\rm co}$, $\Delta k$, $\varepsilon_{\rm 4,s}-\varepsilon_{\rm F}$, and $\varepsilon_{\rm F}-\varepsilon_{\rm 3,s}$ as a function of $P$, respectively. 
}
\label{fig_10_3_2}
\end{figure}

By utilizing $\nu=3/4$, we can determine $\mu$ through 
\begin{equation}
\nu =\frac{1}{4N_k} \sum_{i=1}^{4} \sum_{{\bf k}}
f(\varepsilon(i,{\bf k})).
\label{mu_N}
\end{equation}
Fig. \ref{fig_8} shows the weak $T$-dependence of $\mu$, which arises from the slight band asymmetry. In Fig. \ref{fig_10}, we present the electronic specific heat, $C$, obtained from numerical calculations at $P/P_{\rm merge}=0.511$ and 1.000. These results can be effectively fitted by $C\propto T^{1.9995}\simeq T^{2.0}$ and  $C\propto T^{1.5029}\simeq T^{1.5}$, respectively, within the low temperatures range ($0.9$ K $\leq T\leq$ 1.3 K). At $P=39.4$, as depicted in Fig. \ref{fig_10}, $C$ diminishes significantly for $T\lesssim 3.0$ K due to the presence of an energy gap.


In the regime where the merging of two Dirac points is imminent ($0.894\leq P/P_{\rm merge}\leq 0.996$), the $T$-dependence of $C$ exhibits a crossover: $C\propto T^{2.0}$ at low temperatures and $C\propto T^{1.5}$ at high temperatures [Figs. \ref{fig_10_3} (a) and (b)]. This crossover is confirmed by the nearly constant behavior observed at low temperatures in the $T$-dependencies of $C/(k_{\rm B}T^{2.0})$ and the nearly $1/\sqrt{T}$ dependence at high temperatures, as shown in Figs. \ref{fig_10_4}. 
Similar to Section \ref{honey}, we define the crossover temperature, $T_{\rm co}$, using a maximum in $C/(k_{\rm B}T^{2.0})$. 
The $P$-dependences of $T_{\rm co}$, $\varepsilon_{\rm 4,s}-\varepsilon_{\rm F}$, $\varepsilon_{\rm F}-\varepsilon_{\rm 3,s}$, and $\Delta k$ in Fig. \ref{fig_10_3_2} are shown in Fig. \ref{fig13_2}, resembling the $t_a/t$-dependences in the honeycomb lattice. The $P$-dependence of $T_{\rm co}$ can be approximately described by
\begin{eqnarray} 
T_{\rm co}=-110(P/P_{\rm merge}-1.000) {\rm K}, \label{Tco_B}
\end{eqnarray} 
represented by the green line in Fig. \ref{fig_10_3_2}. 

We ignore the interlayer hopping, $t_{\perp}$, of $\alpha$-(BEDT-TTF)$_2$I$_3$. By introducing $t_{\perp}$, the quasi-two-dimensionality of the system becomes a crucial factor for the electronic specific heats at very low temperatures, because the density of states near the Fermi energy is changed. Consequently, the determination of the crossover becomes ambiguous at extremely low temperatures. Moreover, recent proposals have suggested the existence of a three-dimensional Dirac fermion at very low temperatures\cite{morinari2020,tajima2023}. If this is the case, the power law behavior for $C$ would follow $C\propto T^{3.0}$, which starkly contrasts with the behavior, $C\propto T^{2.0}$, observed in two-dimensional Dirac fermion systems. Hence, further investigation is required to elucidate the electronic specific heats in quasi-two-dimensional Dirac fermion systems, particularly at very low temperatures.

\section{Conclusions}

In this study, we performed comprehensive numerical calculations to explore the intriguing behavior of the electronic specific heat, $C$, in two-dimensional systems employing simplified models for the honeycomb lattice 
and $\alpha$-(BEDT-TTF)$_2$I$_3$. Our findings shed light on the behavior of $C$ when two Dirac points approach each other. Specifically, we observed a transition from $C\propto T^{2.0}$ at low temperatures to $C\propto T^{1.5}$ above the crossover temperature, $T_{\rm co}$, which we defined based on the maximum in $C/T^{2.0}$. Notably, we demonstrated that $T_{\rm co}$ decreases as the spacing between the two Dirac points decreases, eventually reaching zero when the points merge. This merging process is associated with a topological phase transition. Thus, measuring $T_{\rm co}$ provides a promising method to investigate the motion and merging of two Dirac points in two-dimensional Dirac fermion systems. The occurrence of the crossover serves as a significant indicator of the topological phase transition. Additionally, in the $T-C/T^{2.0}$ plot, we anticipate observing a maximum attributed to the presence of the saddle point.

For the honeycomb lattice with only the nearest neighbor transfer integrals ($t_a$, $t_b$, and $t_c$), the semi-Dirac point emerges at $t_a/t=2.00$\cite{Dietl2008,Hasegawa2006}. Therefore, if $t_a/t\simeq 2.00$ can be achieved and the Coulomb interaction can be neglected, $T_{\rm co}$ will be observed in graphene and artificial systems with the honeycomb lattice.

Regarding $\alpha$-(BEDT-TTF)$_2$I$_3$, the crossover is expected to occur near the uniaxial strain of $39.155$ kbar when the interpolation formula\cite{Katayama2004,Katayama2006,Suzumura2013} is used. The experiment for electronic specific heat\cite{Konoike2012} has been performed under the hydrostatic pressure of 15 kbar, where the crossover has not been observed yet. The crossover may be observed by the experiments at higher hydrostatic pressures. In another quasi-two-dimensional massless Dirac fermion, $\alpha$-(BETS)$_2$I$_3$, the crossover is also expected to occur if the situation where two Dirac points are nearly merged can be realized, for example, through the application of uniaxial pressure.

\newpage


\begin{thebibliography}{99}



\bibitem{novo2005}
K. S. Novoselov, A. K. Geim, S. V. Morozov, D. Jiang, M. I. Katsnelson, I. V. Grigorieva, S. V. Dubonos, and A. A. Firsov, Two-dimensional gas of massless Dirac fermions in graphene, Nature {\bf 438}, 197 (2005).

\bibitem{review} T. Ishiguro, K. Yamaji, and G. Saito, 
Organic Superconductors, 2nd ed., 
(Springer-Verlag, Berlin, 1998).

\bibitem{review2} A. G. Lebed, editor, The Physics of Organic Superconductors and Conductors (Springer, Berlin, 2008).


\bibitem{Katayama2006}
S. Katayama, A. Kobayashi and Y. Suzumura, Pressure-Induced Zero-Gap Semiconducting State in Organic Conductor $\alpha$-(BEDT-TTF)$_2$I$_3$ Salt, J. Phys. Soc. Jpn. 
\textbf{75}, 054705 (2006). 




\bibitem{Kino}
H. Kino and T. Miyazaki, First-Principles Study of Electronic Structure in  $\alpha$-(BEDT-TTF)$_2$I$_3$ at Ambient Pressure and with Uniaxial Strain, J. Phys. Soc. Jpn. {\bf 75}, 034704 (2006).



\bibitem{Alemany2012}
P. Alemany, J.P. Pouget and E. Canadell, Essential role of anions in the charge ordering transition of 
$\alpha$-(BEDT-TTF)$_2$I$_3$, Phys. Rev. B \textbf{85}, 195118 (2012).


\bibitem{Kajita2014}
K. Kajita, Y. Nishio, N. Tajima, Y. Suzumura and 
A. Kobayashi, Molecular Dirac Fermion Systems -Theoretical and Experimental Approaches-, 
J. Phys. Soc. Jpn. \textbf{83}, 072002 (2014).




\bibitem{Osada2008}
T. Osada, Negative Interlayer Magnetoresistance and Zero-Mode Landau Level
in Multilayer Dirac Electron System, J. Phys. Soc. Jpn. \textbf{77}, 084711 (2008).

\bibitem{Hirata2011}
M. Hirata, K. Ishikawa, K. Miyagawa, K. Kanoda and M. Tamura, $^{13}$C NMR study on the charge-disproportionated conducting state in the quasi-two-dimensional organic conductor $\alpha$-(BEDT-TTF)$_2$I$_3$, Phys. Rev. B {\bf 84}, 125133 (2011).

\bibitem{Konoike2012}
T. Konoike. K. Uchida and T. Osada, Specific Heat of the Multilayered Massless Dirac Fermion System, J. Phys. Soc. Jpn. {\bf 81}, 043601 (2012).

\bibitem{Ohki2020} 
D. Ohki, K. Yoshimi, and A. Kobayashi, Transport properties of the organic Dirac electron system $\alpha$-(BEDT-TTF)$_2$I$_3$, Phys. Rev. B {\bf 102}, 235116 (2020).


\bibitem{Tsumu2021} 
T. Tsumuraya and Y. Suzumura, First-principles study of the effective Hamiltonian for Dirac fermions 
with spin-orbit coupling in two-dimensional molecular conductor $\alpha$-(BETS)$_2$I$_3$, Eur. Phys. J. B {\bf 94}, 17 (2021).


\bibitem{kitou2021} 
S. Kitou, T. Tsumuraya, H. Sawahata, F. Ishii, K. I. Hiraki,
T. Nakamura, N. Katayama, and H. Sawa, Ambient-pressure Dirac electron system in the quasi-two-dimensional molecular conductor $\alpha$-(BEDT-TTF)$_2$I$_3$, Phys. Rev. B {\bf 103}, 035135 (2021).




\bibitem{Tajima2021} 
Y. Kawasugi, H. Masuda, M. Uebe, H. M. Yamamoto, R. Kato, Y. Nishio, and N. Tajima, Pressure-induced phase switching of Shubnikov-de Haas oscillations in the molecular Dirac fermion system $\alpha$-(BETS)$_2$I$_3$, Phys. Rev. B {\bf 103}, 205140 (2021).




\bibitem{morinari2020} 
T. Morinari, Dynamical Time-Reversal and Inversion Symmetry Breaking, Dimensional Crossover, and Chiral Anomaly in  $\alpha$-(BEDT-TTF)$_2$I$_3$, J. Phys. Soc. Jpn. {\bf 89}, 073705 (2020).



\bibitem{tajima2023} 
N. Tajima, Y. Kawasugi, T. Morinari, R. Oka, T. Naito, and R. Kato, Coherent Interlayer Coupling in Quasi-Two-Dimensional Dirac Fermions in  $\alpha$-(BEDT-TTF)$_2$I$_3$, J. Phys. Soc. Jpn. {\bf 92}, 013702 (2023).



\bibitem{Hasegawa2006}
Y. Hasegawa, R. Konno, H. Nakano, and M. Kohmoto, Zero modes of tight-binding electrons on the honeycomb lattice, Phys. Rev. B \textbf{74}, (2006) 033413.


\bibitem{Dietl2008} 
P. Dietl, F. Piechon, and G. Montambaux, New Magnetic Field Dependence of Landau Levels in a Graphenelike Structure, Phys. Rev. Lett. {\bf 100}, 236405 (2008).


\bibitem{Bane} 
S. Banerjee, R. R. P. Singh, V. Pardo, and W. E. Pickett, Tight-Binding Modeling and Low-Energy Behavior of the Semi-Dirac Point, Phys. Rev. Lett. {\bf 103}, 016402 (2009).




\bibitem{Suzumura2013}
Y. Suzumura, T. Morinari and F. Piechon, Mechanism of Dirac Point in $\alpha$ Type Organic Conductor under Pressure, J. Phys. Soc. Jpn. \textbf{82} 023708 (2013).




\bibitem{Tarruell2012} 
L. Tarruell, Daniel Greif, Thomas Uehlinger, Gregor Jotzu, and Tilman Esslinger, Creating, moving and merging Dirac points with a Fermi gas in a tunable honeycomb lattice, Nature {\bf 483}, 302 (2012).



\bibitem{Bellec} 
M. Bellec, U. Kuhl, G. Montambaux, and F. Mortessagne, Topological Transition of Dirac Points in a Microwave Experiment, Phys. Rev. Lett. {\bf 110}, 033902 (2013).


\bibitem{Mil} 
M. Mili\'{c}evi\'{c}, G. Montambaux, T. Ozawa, O. Jamadi, B. Real, I. Sagnes, A. Lema\^{i}tre, L. Le Gratiet, A. Harouri, J. Bloch, and A. Amo, Type-III and Tilted Dirac Cones Emerging from Flat Bands in Photonic Orbital Graphene, Phys. Rev. X {\bf 9}, 031010 (2019).


\bibitem{real} 
B. Real, O. Jamadi, M. Mili\'{c}evi\'{c}, N. Pernet, P. St-Jean, T. Ozawa, G. Montambaux, I. Sagnes, A. Lema\^{i}tre, L. Le Gratiet, A. Harouri, S. Ravets, J. Bloch, and A. Amo, Semi-Dirac Transport and Anisotropic Localization in Polariton Honeycomb Lattices, Phys. Rev. Lett. {\bf 125}, 186601 (2020).







\bibitem{kittel}
Charles Kittel, Introduction to Solid State Physics, 8th ed., 
(John Wiley \& Sons, Inc. 2005). 



\bibitem{Vaf}
 O. Vafek, Anomalous Thermodynamics of Coulomb-Interacting Massless Dirac Fermions
in Two Spatial Dimensions, Phys. Rev. Lett. {\bf 98} (2007) 216401.


\bibitem{She}
D. E. Sheehy and J. Schmalian, Quantum Critical Scaling in Graphene, Phys. Rev. Lett. {\bf 99} (2007) 226803.



\bibitem{Kot}
V. N. Kotov, B. Uchoa, V. M. Pereira, F. Guinea, and A. H. Castro Neto, Electron-Electron Interactions in Graphene: Current Status and Perspectives, Rev. Mod. Phys. {\bf 84}, 1067 (2012).





\bibitem{wallace}
P. R. Wallace, The Band Theory of Graphite, Phys. Rev. {\bf 71}, (1947) 622. 



\bibitem{Reich}
S. Reich, J. Maultzsch, C. Thomsen, and P. Ordejon, Tight-binding description of graphene, Phys. Rev. B {\bf 66}, 035412 (2002).



\bibitem{KH2017}
K. Kishigi, and Y. Hasegawa, Three-quarter Dirac points, Landau levels, and magnetization in $\alpha$-(BEDT-TTF)$_2$I$_3$, Phys. Rev. B {\bf 96}, 085430 (2017).


\bibitem{Katayama2004}
A. Kobayashi, S. Katayama, K.Noguchi and Y. Suzumura, Superconductivity in Charge Ordered Organic Conductor -$\alpha$-(ET)$_2$I$_3$ Salt-, 
J. Phys. Soc. Jpn. 
\textbf{73}, 3135 (2004). 






\bibitem{Mori1984}
T. Mori, A. Kobayashi, Y. Sasaki, H. Kobayashi, G. Saito, and
H. Inokuchi, BAND STRUCTURES OF TWO TYPES OF (BEDT-TTF)$_2$I$_3$, Chem. Lett. \textbf{13}, 957 (1984).



\bibitem{Kondo2005}
R. Kondo, S. Kagoshima, and J. Harada,Crystal structure analysis under uniaxial strain at low temperature using a unique design of four-axis x-ray diffractometer with a fixed sample, Rev. Sci. Instrum. \textbf{76}, 093902 (2005).













	


































































































































































































































































































\end{thebibliography}


\end{document}